# Analysis of Spatio-Temporal Preferences and Encounter Statistics for DTN Performance

Gautam S. Thakur*, Udayan Kumar*, Ahmed Helmy*, Wei-Jen Hsu+
*Department of Computer and Information Science and Engineering, University of Florida, Gainesville, FL
+Cisco Systems Inc., San Jose, CA
*{gsthakur, ukumar, helmy}@cise.ufl.edu, +wehsu@cisco.com

*Abstract*—Spatio-temporal preferences and encounter statistics provide realistic measures to understand mobile user's behavioral preferences and transfer opportunities in Delay Tolerant Networks (DTNs). The time dependent behavior and periodic reappearances at specific locations can approximate future online presence while encounter statistics can aid to forward the routing decisions. It is theoretically shown that such characteristics heavily affect the performance of routing protocols. Therefore, mobility models demonstrating such characteristics are also expected to show identical routing performance. However, we argue models despite capturing these properties deviate from their expected routing performance. We use realistic traces to validate this observation on two mobility models. Our empirical results for epidemic routing show those models' largely differ *(delay 67% & reachability 79%)* from the observed values. This in-turn call for two important activities: (i) Analogous to routing, explore structural properties on a Global scale (ii) Design new mobility models that capture them.

## I. INTRODUCTION

Over the past decade, popularity of mobility models has increased significantly. They are inexpensive and versatile tools to measure network performance. Much of the recent work in modeling space investigate communication paradigm for extreme environments [14], where infrastructure support is inadequate, long delays and frequent network partitioning characterize the transmission. Considered a necessary evil, mobility further exacerbates the problem of severely power and memory constraint mobile nodes. Invariably, all these issues make an extraordinary impact on the design and development of new routing protocols, topologies and network tools.

Recent advancements to understand such settings come from human carrying multi-sensor devices (like laptops, iPhones, etc.). Their tight coupling gives a way to capture essential features of mobility that we can use for modeling purposes. For example, a combination of Bluetooth and WiFi access point sniffers yields remarkable information about mobile users' spatio-temporal presence and interaction with others. The criterion is even extended to vehicular [4] and inter-planetary networks [6]; information diffusion in disaster relief networks [38]. Furthermore, study of these sensory inputs using social and network theory derive interesting facts about human behavioral patterns and their association to other users and locations. Therefore, mobility models demonstrating such patterns are a great way to measure the network performance. We expect models reproducing identical patterns also deliver statistically equivalent network performance result for message delivery, delay overhead etc. Without any exception, they should also capture the underlying structural and dynamical properties. Here we seek to addresses all these issues. We study few models some of them are driven by random mobility while other caters specific scenarios.

In this paper, first we analyze spatio-temporal properties [22] and encounter statistics on two different realistic wireless measurements. Later on we evaluate the same characteristic on the synthetic traces produced by two different mobility models. We study these two encounters statistics: 1) Inter-meeting Time. 2) Meeting Duration. Finally, we perform epidemic routing [40] on the realistic scenario and on the synthetic traces to compare their network performance. Surprisingly, in doing so, we find that despite mobility models reflect equivalent spatio-temporal and encounter statistics, they completely miss out on the routing performance metrics. The results of epidemic routing show mobility models deviate 79% on average reachability, 67% on the delay and 58% on the overheads as compared to realistic scenarios. These are surprising to us and in a way, an opportunity to re-visit the design and development of mobility modeling space. In our on-going study, we are also investigating whether global characteristics like network density, mobility coefficient, clustering coefficient affect the overall networking performance.

In the ensuing text, section II illustrate related work; details of the datasets and trace analysis are explained in section III. We devote section IV to briefly describe our two mobility models, then in section V we explain studied characteristic and evaluate the results. Finally, we conclude our paper in section VI by giving future work direction.

## II. RELATED WORK

The casual meetings (also known as encounters) among participating entities [14] drive the communication in opportunistic settings like DTNs. However, this simplicity poses serious challenges [44] to design new routing protocols, mobility models and among others also establish topology constraints in estimating network performance. For example, in DTNs one cannot guarantee pre-established communication links, infrastructure support, and predictability to receive and acknowledge a message. These challenges make researcher to think from a very different perspective - why not to investigate into the state-space and characteristics of agents involved in the communication. For example, in human mobility settings, understanding their behavioral patterns can aid in predicting the future transmissions and network reachability. On these lines authors in [8] study the impact of human mobility in making forwarding decisions. They found inter-meeting time between mobile entities follow a power-law distribution over a large set of values. Contrary to that, in [25] authors discover that inter-meeting time follow power-law up-to a characteristic time beyond which its decay is exponential. In [45], authors analyze hitting time for a number of mobility models and derive expression to estimate their accurate values. Recently, the trends in gathering mobility measurements from real world [29, 39, 43] also motivate to investigate patterns in regularity [12], periodicity [21, 26], community behavior [13, 16, 34, 37, 17] and in general behavioral patterns [5, 10, 20, 24, 28, 33] of mobile users to a large extent. In all, these statistical studies can prove important into the development of new DTN mobility models.

Table 1:Details of the Wireless Measurements

| Location | # Users | Duration |
|---|---|---|
| Cambridge Infocom | 41 | 3-4 days |
| MIT Campus | 1366 | Fall 2006 |

While synthetic mobility models exhibit specific scenarios and give close form of expressions to bound performance, trace driven models focus more on the realistic scenarios and give an insight into the qualitative characteristics of mobility. Sometimes they are also mathematically tractable. In [18, 37], authors propose group mobility models to understand the community based social interaction in ad hoc networks. Motivated by the realistic scenarios work in [27, 31, 47] illustrate step-by-step process to extract mobility models from WLAN measurements. In [30, 41] authors show that human mobile users perform a truncated levy walk and give estimations for online and pause time. They also discover, a heavy tail distribution on human walks is because of the similar and burst preferential attachments to the locations. An in-depth study of statistical properties on the frequent transmission impairments, sparse distribution of mobile users, mobility-assisted routing and metrics in [1, 8, 11, 15, 40] provide good reasoning about developing new mobility models for the DTN scenarios. Our study is motivated by [22] that capture the non-homogeneous behaviors of mobile users in both space and time. The model generate measurements to show that display (i) skewed location visiting preferences; (ii) time dependent periodical reappearance of mobile users as seen in WLAN measurements along with other encounter statistics like average node degree, hitting and meeting time. Interested reader can always refer to mobility models analysis done in [2, 3, 7].

Finally, in DTNs an underlying constraint in effective communication is the inability to pre-establish a complete route from source to destinations. Thus popular ad hoc routing protocols like AODV and DSR fail to work properly. Keeping with the spirit of network dynamics, several DTN protocols are available that provide a good component to better route messages in the opportunistic network settings. Some of those studies include [9, 19, 23, 35, 36] that provide way to capture the delay associated with routing decisions as well as take the advantage of social forwarding to enhance and better predict routing decisions. To baseline the results authors in [32, 46, 40] proposed flooding, replication-based low resource utilization and history-based protocols to independently route with any network topology settings. In [19] authors propose *Profile-Cast*, a behavior oriented communication paradigm for human mobile networks. The author's longitudinal analysis on spatio-temporal behavioral profiles show enough stability that helps to achieve delivery rates very close to the delay-optimal strategy with minimal overhead.

Table 2: Anonymized Encounter Sample of Mobile User WLAN session

| Node Mac ID | Location | Start Time | End Time |
|---|---|---|---|
| aa:bb:cc:dd:ee:ff | Loc-1 | 44400343 | 76404567 |
| a1:b2:c3:d4:e5:f6 | Loc-1 | 64300343 | 86895742 |
| a7:b8:c9:d1:e2:f3 | Loc-1 | 56744343 | 89404567 |
| a4:b5:c6:d7:e8:f9 | Loc-4 | 62846767 | 88878766 |

For our study, we mainly focus the mobility modeling and routing aspects in the delay tolerant networks.

### III. DATA SET AND TRACE ANALYSIS

#### A. Data Set

In order to perform any system study, it is important to know beforehand, the investigation space and the nature of the data that is used to support the hypothesis. A preliminary analysis on the dataset must conform a require level of granularity and the existence of patterns that we expect to study. In our experiment, we examine two types of data sets. First, we make use of publicly available WLAN session measurements of MIT campus consisting of 1300+ active users over a period of one month. On this dataset we perform a study to understand the preferential attachment to certain locations and time-dependent periodic behavior of mobile users. Our second dataset comprises of Bluetooth encounter traces from IEEE Infocom 2005 iMotes experiment. In this, Intel's iMote were distributed to 41 participants for a period of three-four days. On this dataset we investigate the encounter statistics and construct mobility model that captures these characteristics. Table-1 gives the detail of these measurements.

#### B. Encounter Traces

In order to pursue the study on the encounter statistics and dynamic routing in DTN, we need measurements that quantitatively depict the meeting (*a.k.a. encounter*) between mobile users. An encounter occurs when their devices come in the radio communication range. This is straight forward for the Bluetooth measurements that comes with precise encounter information. However, the WLAN measurements are accumulated at the access point level and contain usage patterns. Luckily, we have the advantage to convert these measurements into user encounter patterns. *We consider encounter in WLAN if two users are connected under the same access point and their online session time is a non-empty set*. However, a counter argument can be established, but WLAN measurements have the advantage to obtain traces in much larger sizes with richer user presence. This argument can be tested from Table-1, where experimental approaches involves distribution of multi-sensor devices for a small user base and limited time period. In Table-2 we show a sample of anonymous encounter trace. Note that last row user (*Mac address*) do not encounter with others because its online location is not the same despite intersecting session times.

### IV. MOBILITY MODELS STUDIED

In this section, we discuss two mobility models used to evaluate mobility characteristics. We use Random Direction Model [42] to establish lower bound estimates on the quantitative performance using its random mobility characteristics. On other side, wide applicability of trace driven mobility models give a blueprint of realistic scenarios and statistically accurate results. These features attract us to use Time Variant Community model [22]. For the purpose of our study, we also find that it best captures spatio-temporal preferences and encounter statistics. In the following text, we briefly describe these models and construct trace driven DTN scenarios to estimate routing performance.

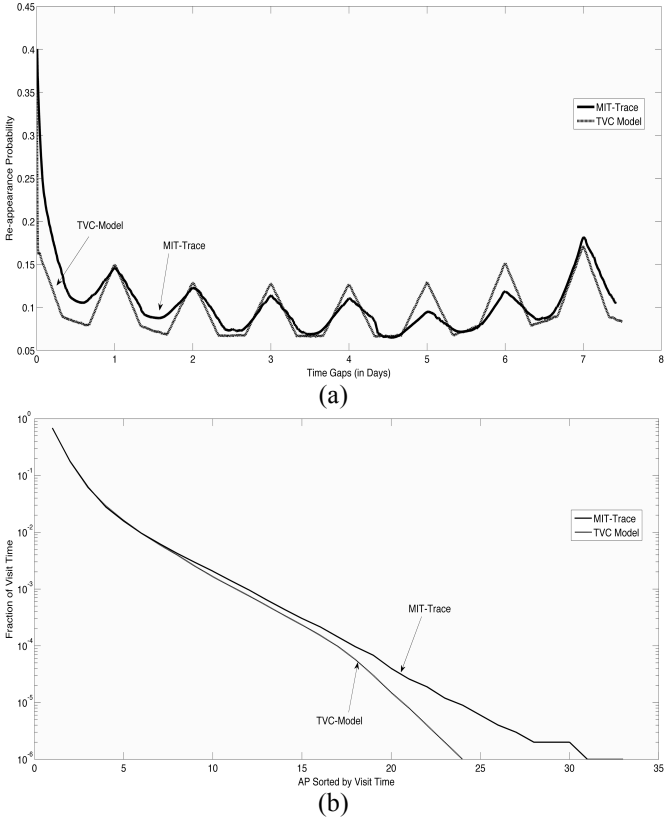
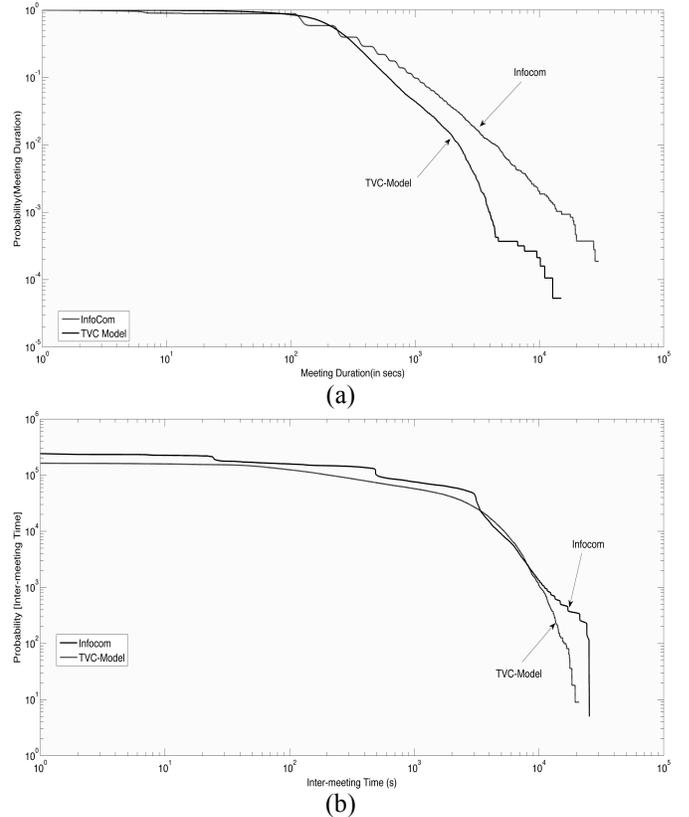

**Figure 1:** (a) Periodic re-appearances of mobile users for the MIT Campus. TVC Model recreates similar preferences as observed in real traces. (b) TVC depicts matching skewed location visiting preferences as observed in real MIT campus traces.

**Figure 2:** (a) TVC Model depicts meeting duration as measured in the real Infocom traces. (b) Inter-meeting time between mobile users are also similar for the TVC and real Infocom traces.

### A. Random Direction Mobility Model

In random direction model, a mobile node randomly selects a movement degree to travel in a particular direction until it reaches the destination boundary area with a given speed. On reaching, it stops for a given pause time before selecting a new direction to move. This model is more stable as compared to other random models and provides quantitatively even distribution of nodes in the simulation area. We setup this model to investigate the effect of random movements on DTN performance. We modify this model in two ways: in one setting we introduce on/off behavior of mobile nodes along with random pause times, which is equivalent to the realistic online/offline nature of mobile nodes. In other, we keep standard settings to baseline comparisons with pure form of randomness.

### B. Time Variant Community Mobility Model

The TVC model [22] depicts two realistic mobility features, which are very prominent in wireless networks and detected in real traces: (1) skewed location visiting preferences; (2) periodic re-appearances at the locations. For mobile nodes often visits, TVC model propose the concept of preferential attachment to geographical locations, or the *communities*, in order to observe that mobile users spend significant portion of their online time at a selected few locations. Furthermore, the TVC model also introduces time periods, which allows duration based mobility preferences for users in a recurring interval fashion. The best part of TVC model is the capacity to realistically reproduce user behavioral preferences and the versatility to fine-tune for different mobility scenarios. By using TVC model, we implicitly evaluate the results and properties as reflected by models in [42, 45]. We configure the TVC for two different settings: (i) to model identical spatio-temporal preferences visualized in real MIT measurements (ii) to demonstrate identical encounter statistics seen in Infocom iMote measurements. Later on, we use the synthetic TVC traces of these settings to evaluate routing performance.

### V. EVALUATION OF MOBLITY CHARACTERISTICS

In this section, first we provide a brief overview of spatio-temporal preferences and encounter statistics. Later on, we construct mobility models to demonstrate them via synthetic traces. Finally, we run epidemic routing on real and synthetic measurements and analyze the results.

### A. Evaluation Metrics

#### 1) Analysis of Spatio Temporal Preferences

We capture two types of non-homogenous behavior of mobile users in both space and time. They are: (i) Skewed location visiting preferences (ii) Periodical reappearances. Studies carried out in [12, 20-22, 26] tell us that mobile user exhibit preferential attachment to few locations and behavioral regularity in DTNs. We assume, understanding these distributions aid to better

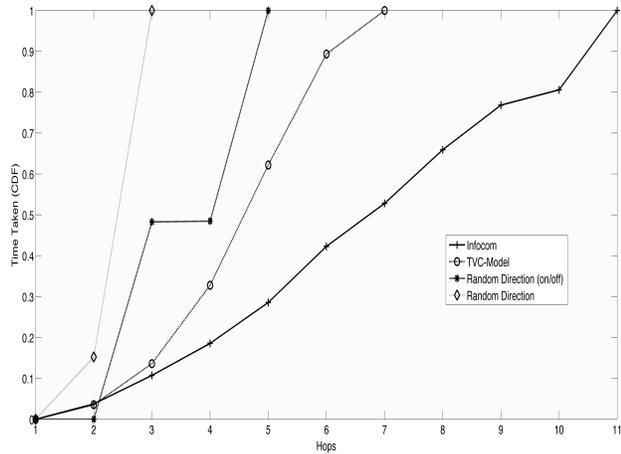

(a) Delay [Infocom]

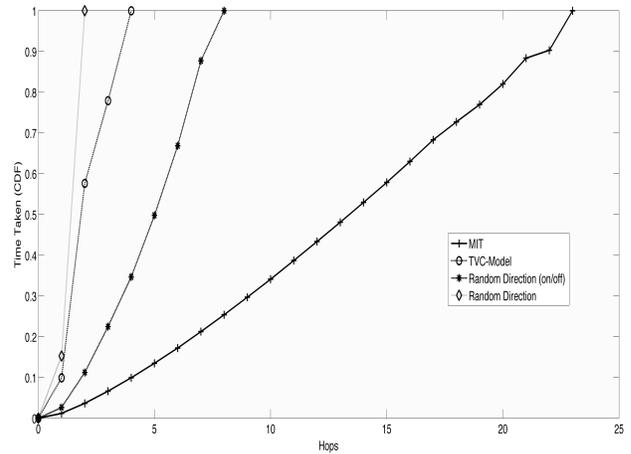

(d) Delay [MIT Campus]

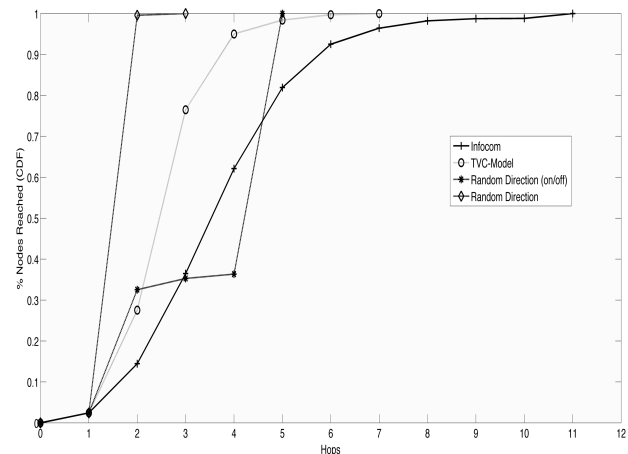

(b) Reachability [Infocom]

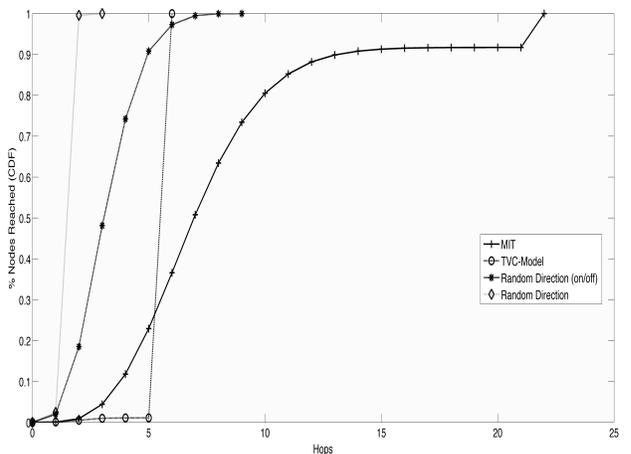

(e) Reachability [MIT Campus]

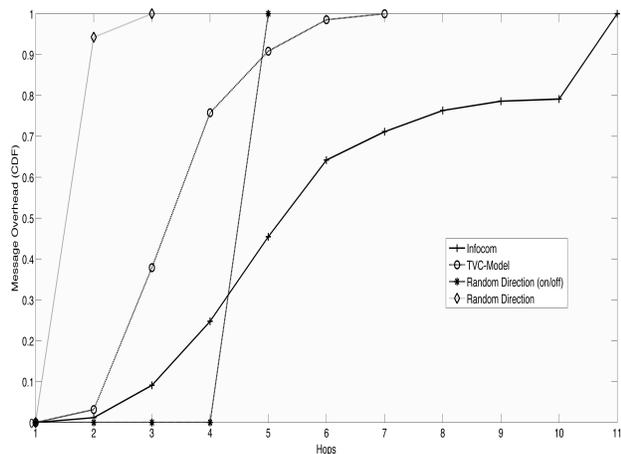

(c) Overhead [Infocom]

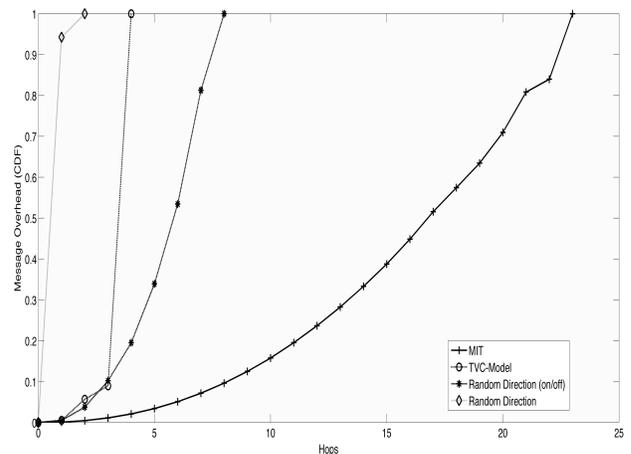

(f) Overhead [MIT Campus]

**Figure 3: (a-c) Show Epidemic routing results for the Infocom settings. (d-f) Show Epidemic routing results for the MIT settings. As seen, both TVC and random direction model largely deviate in their network performance for delay, overhead and message delivery compared to real measurement results.**

message dissemination, prediction of information transmission and the message delivery in opportunistic setting.

*2) Analysis of Encounter Statistics*

In dynamic infrastructure-less mobile networks (like DTNs etc.), the routing is performed by data carrying mobile nodes. The

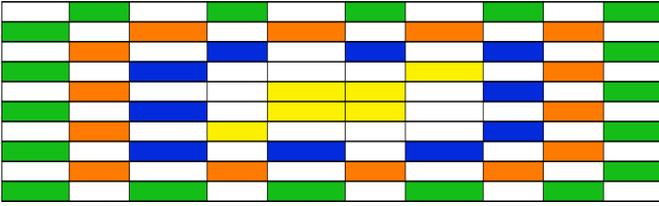
Figure 4: Conference and hotel room settings for Infocom trace in 100x100 grid. In center, the yellow color filled boxes are conference settings while other three colors show individual hotel rooms of the 41 participants.

exchange of information takes place when two nodes encounter (a.k.a meet) each other. Intuitively, we can improve routing mechanism given we understand the statistics of these encounter patterns. So, we analyze two encounter statistics: (i) Intermeeting Time, which is the time gap that separates two consecutive mobile encounters. (ii) Meeting Duration, which is the single uninterrupted meeting session time surrounded by intermeeting distribution. Thus, our statistics alternate between each other.

*3) DTN Performance Metrics*

We look into delay, reachability and message overhead to estimate epidemic routing performance. First, we convert the synthetic usage traces from the models into encounter traces. The encounter traces shown in Table-2, is time varying intermittently connection opportunity sessions between mobile nodes. Then we use the implicit time scale of traces to model the network dynamics and encounters as a medium to exchange epidemic routing information. Finally, we run epidemic routing on the settings to compare the performance of Random Direction, TVC and real measurements. Formally, *Reachability* is number of hops it took for source node's messages to reach all the recipients; *Delay* is the total time taken by source node's messages to reach all the recipients; *Overhead* is the average message count percentage incurred during the simulation.

*B. Results*

Here, we investigate how synthetic measurements produced by the mobility models deviate from realistic consideration. Initially, we construct TVC model to imitate spatio-temporal preferences observed in the realistic mobility patterns of MIT campus. Then we re-configure the model to demonstrate identical encounter statistics of Infocom iMotes experiment. Since random direction is not designed to model these characteristics, we skip its evaluation and instead use its traces to focus on the routing performance only. Then we take combined synthetic traces earlier generated by the TVC (for MIT and Infocom) and random direction model to evaluate the epidemic routing performance against the respective set of real traces.

*1) Spatio-temporal Analysis for MIT Campus*

We construct the TVC model to generate a month long synthetic trace for 1366 nodes. The statistical analyses of the results are plotted in Fig.-1. As expected, TVC successfully replicate realistically similar location visiting and periodic reappearances properties for the MIT. Here on we conclude that TVC provide true measures to capture non-homogenous behavioral patterns of mobile users and demonstrate their spatio-temporal preferences.

*2) Encounter Statistic Analysis for Infocom*

Next, we construct TVC model to generate synthetic traces that imitate identical Infocom encounter statistics. We model individual behavior patterns of 41 nodes for four days in a conference area like setting with flexibility to visit hotel rooms and outside locations. Fig.-4 shows this setting on a 100x100 grid area. The CDF plots of Fig.-2 show TVC captures identical encounter statistics observed by the experiment. We see intermeeting time follows powerlaw distribution up to a characteristic time period after which it decay exponentially. This clearly shows TVC can be used for statistical analysis of encounter patterns for unknown scenarios also.

*3) Analysis of Epidemic Routing*

We implement epidemic routing in C++ that input time varying mobile encounter sessions. Essentially, they serve a basis to model intermittently connected dynamic network topology setting where each encounter is viewed as an opportunity to receive and forward messages. The CDF plots of Fig-3 show the routing performance results for both MIT and Infocom, and quantitative report in the Table-3. Surprisingly in both the cases, despite mobility models claim to exhibit previous characteristics, they dramatically deviate in network routing performance benchmarks. We observe for the reachability in Infocom setting, an average node takes 11 hops to deliver messages to all other nodes; while TVC takes only 7 and random direction models take even less. Similar reachability trends are seen for MIT campus traces also. Meanwhile for the delay, there are at least two folds of difference between realistic and synthetic measurements. The TVC and the random direction models take less time in delivering messages compared to the forwarding patterns observed in the realistic scenario. Finally, TVC generated message count overhead also diverge from the realistic values.

*C. Results Discussion*

The in-depth analysis show models capture spatio-temporal preferences and encounter statistics, but completely miss out on the performance evaluation criteria for routing in disconnected networks. However, we iterate these mobility models surely serve the purpose for which they are designed. However looking forward, we believe that scientists not only need models to imitate realistic mobility scenarios, but also identical performance and the depth of underlying structural and topological realisms. A good research direction would be look into measures that affect globally, similar in the way routing decision are made. In turn, we also think it's important to revisit the design aspects of mobility models for a statistically acceptable performance gains.

## VI. CONCLUSION AND FUTURE WORK

In this paper, we scrutinize mobility models on routing performance benchmarks. We testify that despite models capture realistic human behavioral patterns; their routing performance is orthogonal to the trends seen in reality. Initially, we construct the Time Variant Community model to generate synthetic trace, which exhibit identical spatio-temporal preferences and encounter statistics seen in two realistic measurements. Later on, we transform these session traces into intermittently connected encountered traces and also use random direction model to perform epidemic routing. By doing so, we find that mobility models performance is not analogous to realistic trends. The average number of hops and delay it takes to reach all nodes is respectively 57% and 79% less for Infocom and 78% and 80% less for MIT compared to the real trace overheads. These dramatic deviations from realism indicate serious flaws in the existing models and their inadequacy as testbed tools for any kind of performance evaluation purposes. In our on-going and future work, we are looking into a global perspective of clustering and mobility coefficient, density distribution and discovering underlying structural properties and community dynamics. We are also exploring spatio-temporal similarity that can aid in making better routing decisions and planning to re-visit mobility modeling, which is vital for the evaluation and design of next-generation behavior-aware protocols.

Table 3: Summary of performance measurement for epidemic routing.

| Metric | Infocom | | | | | MIT | | | | |
|---|---|---|---|---|---|---|---|---|---|---|
| | *O* | *T* | *R1* | *R2* | *% Dev* | *O* | *T* | *R1* | *R2* | *% Dev* |
| Reachability | 11 | 7 | 4 | 3 | 57% | 22 | 4 | 7 | 3 | 78% |
| Delay | 89 | 24 | 34 | 0.17 | 79% | 66 | 37 | 2.4 | 0.18 | 80% |
| Overhead | 0.006 | 0.004 | 0.003 | 0.001 | 55% | 0.0012 | 0.0002 | 0.0036 | 0.002 | 61% |
| O = Real Traces; T = TVC Model; R1 = Random Direction (on/off); R2 = Random Direction | | | | | | | | | | |